\documentstyle[aps,prl,floats,graphicx,amssymb]{revtex}
\begin{document}
\draft
\title{Hydrostatic theory of superfluid $^3$He-B
\thanks{Associated internet page: 
http://boojum.hut.fi/research/theory/btex.html}}

\author{E.V. Thuneberg} 
\address{Low Temperature Laboratory, Helsinki University of 
Technology, 02150 Espoo, Finland}
\date{\today}
\maketitle
\def\imagu{{\rm i}}

\begin{abstract}  The determination of the texture of the order
parameter is important for understanding many experiments in superfluid
$^3$He. In addition to reviewing the theory of textures in superfluid
$^3$He-B we give several new results, in particular on the surface
parameters in the Ginzburg-Landau region and bulk parameters at
arbitrary temperature.  Special attention is paid to separate the
results that are valid at all temperatures from those which are limited
to the Ginzburg-Landau region. We study the validity of a trivial
strong-coupling model, where the energy gap of the weak-coupling
theory is scaled by a temperature dependent factor. We compare
the theory with several experiments. For some
quantities the theory seems to work fine and we extract the
dipole-dipole interaction parameter from the measurements. 
\end{abstract} 

\section{Introduction} 

The superfluid phases of liquid $^3$He show complex behavior, which
still can be understood theoretically. Many phenomena have been studied
in a pure form in $^3$He, and the knowledge can then be applied
to other physical systems. For example, several structures of quantized
vorticity have been seen in both A and B phases of $^3$He
\cite{LounasT}. The effect of impurities has recently been studied in
many laboratories by aerogel immersed in liquid 
$^3$He \cite{aerogel}. Recent experiments on the Josephson effect
show unexpected behavior \cite{packard2}. Theoretical
understanding of all these phenomena requires good quantitative
understanding of the basic properties of superfluid $^3$He. This
provides the motivation for the present paper. 

The purpose of hydrodynamics is to determine the behavior of a fluid on
length and time scales that are long compared to some microscopic
lengths and times \cite{LLfluid}. Hydrostatics is a subfield of
hydrodynamics. It is restricted to study the equilibrium properties of
the fluid. In simple fluids hydrostatics reduces to statements about
the pressure variation in the fluid, which either rotates uniformly or
is exposed to some external field. The problem becomes more difficult,
if the fluid has some broken symmetry. Particular examples of these are
liquid crystals and superfluids $^4$He and $^3$He. In both superfluids,
the equilibrium mass current belongs to the scope of the hydrostatic
theory. The order parameter of $^3$He has also other degrees of
freedom. The structure of those, which is often called {\it texture},
also has to be incorporated. In $^3$He the hydrostatic theory is
limited to length scales that are large in comparison to the superfluid
coherence length $\xi_0\approx 10\,$nm. The theory is valid at all
temperatures.
The hydrostatic theory can still be applied when the motion of
the quasiparticles at low temperatures becomes ballistic rather than
diffusive.
The hydrostatic theory can be generalized to hydrodynamic theory by
using conservation laws and adding the transport coefficients. 

Our purpose is to make a systematic presentation of the
hydrostatic theory for the B phase of superfluid $^3$He. Part of the
reason is that presently the various results are scattered over a large
literature. Previous reviews treat hydrostatics only as a side topic
and cover only a small part of the subject
\cite{Leggett,BC,SRrev,Fetterrev,SVrev,Kharadze,VW}. The existing
papers are often unclear whether they treat general temperatures or are
restricted to the neighborhood of the superfluid transition
temperature $T_{\rm c}$. We note that the hydrostatics of the A phase is
better presented in the existing literature
\cite{BC,Fetterrev,SVrev,VW} than the B phase considered here. 
In addition to reviewing, we present several results that have
not been published before. 

We will begin with a general formulation of the hydrostatics of
$^3$He-B. Our approach is general enough to allow an external magnetic
field and uniform rotation, both in the leading order. We write down an
energy functional that consists of bulk terms and boundary conditions.
All structures on the length scale of $\xi_0$, such as surface layer or
vortex lines, have to be treated as boundary conditions. The theory is
found to split into two pieces: one for the superfluid velocity and the
other for the texture. The former is identical to the hydrostatics of
superfluid $^4$He, whereas the latter can be solved only after the
superfluid velocity is determined. 

The coefficients of the hydrostatic energy can either be
obtained experimentally or be calculated by some more
microscopic theory. The calculation is discussed in sections
\ref{s.qc} and \ref{s.gl}. The former considers
the quasiclassical theory of $^3$He \cite{SRrev}. The coefficients can
be calculated using the weak-coupling quasiclassical theory  at
arbitrary temperature. We also
discuss a ``trivial strong-coupling''  (TSC) model, where the
weak-coupling coefficients are improved by scaling the energy gap. A
different approach is studied in section 
\ref{s.gl}, where the hydrostatic coefficients are related to the
parameters of the phenomenological Ginzburg-Landau (GL) theory at
$T\approx T_{\rm c}$. 

Before any quantitative tests of the theory, we still have to
determine the parameters that the quasiclassical theory needs as input.
This is discussed in Section \ref{s.gd}, where we analyze several
experiments. We find that certain quantities are reasonably well fitted
using TSC model, but errors of 50\% may occur for other
quantities.  

\section{Hydrostatic free energy}\label{s.f}

The superfluidity in a Fermi system arises from formation of Cooper
pairs. A macroscopic number of pairs occupies the
same pair state in the superfluid state \cite{BCS}. The
relative orbital wave function of a pair has p-wave symmetry in $^3$He
and the spin state is a triplet
\cite{Leggett,WheatleyRev}.  The state of the pairs is thus described by
an order parameter, which is a complex
$3\times 3$ matrix $A_{\alpha i}$. It gives the projections of the
Cooper-pair wave function on the three p-wave orbitals ($p_x$, $p_y$,
and $p_z$, index
$i$) and on the three spin triplet states
($\vert-\uparrow\uparrow+\downarrow\downarrow\rangle$,
${\rm i}\vert\uparrow\uparrow+\downarrow\downarrow\rangle$, and
$\vert\uparrow\downarrow+\downarrow\uparrow\rangle$,
index $\alpha$). In
unperturbed B phase the order parameter has the form
\begin{equation} A_{\alpha j}=\Delta e^{\imagu\phi}R_{\alpha j}
\label{e.2.1}\end{equation} with real $\Delta$, $\phi$ and $R_{\alpha
j}$. Here the amplitude $\Delta$ has a fixed (temperature and pressure
dependent) value and
$R_{\alpha i}$ is constrained to be a rotation matrix, i.e.\
$R_{\alpha i}R_{\alpha j}=\delta_{ij}$. (Summation over repeated
indices is assumed.) The phase
$\phi$ and the more detailed form of the spin-orbit rotation matrix
$R_{\alpha i}$ are not fixed on the scale of the superfluid
condensation energy. These soft variables allow a dissipationless flow
of both mass and spin. The order parameter can be interpreted as the
wave-function of the center of mass of a pair. Using standard quantum
mechanics, we can then define a mass-flow velocity
\begin{equation} {\bf v}_{\rm s}={\hbar\over
2m}{\mbox{\boldmath$\nabla$}}\phi,
\label{e.2.2}\end{equation}
 where $m=5.0097\cdot 10^{-27}$ kg is the mass of a
$^3$He atom. In a similar way one can also define a spin-flow velocity
\begin{equation} {\bf v}^{\rm spin}_{{\rm
s},\alpha}=-{\hbar\over4m}\epsilon_{\alpha\beta\gamma}R_{
\beta i}{\mbox{\boldmath$\nabla$}} R_{\gamma i}
\equiv {\hbar\over 4m}R_{\alpha i}\epsilon_{ijk}R_{\beta
j}{\mbox{\boldmath$\nabla$}} R_{\beta k},
\label{e.2.3}\end{equation} 
 where $\epsilon_{ijk}$ is the maximally
antisymmetric tensor. For example, if the axis of the spin-orbit
rotation is constant and parallel to $z$, the pairs with spin states 
$\vert\uparrow\uparrow\rangle$, 
$\vert\downarrow\downarrow\rangle$, and
$\vert\uparrow\downarrow+\downarrow\uparrow\rangle$
flow with velocities 
${\bf v}_{\rm s}+ {\bf v}^{\rm spin}_{{\rm s},z}$,
${\bf v}_{\rm s}- {\bf v}^{\rm spin}_{{\rm s},z}$, and 
${\bf v}_{\rm s}$,
respectively. The
three-dimensional rotation matrices are conveniently parametrized by an
angle $\theta$ and an axis
$\hat{\bf n}$ of rotation as
\begin{equation} R_{ij}(\hat{\bf
n},\theta)=\cos\theta\delta_{ij}+(1-\cos\theta)\hat n_i\hat
n_j-\sin\theta\epsilon_{ijk}\hat n_k,
\label{e.2.4}\end{equation} where $\hat{\bf n}\cdot\hat{\bf n}=1$. We
note also the trivial identity
$\hat{\bf a}\cdot
\tensor{R}\cdot\hat{\bf a}=\cos\theta+(1-\cos\theta)(\hat{\bf
a}\cdot\hat{\bf n})^2$, where $\hat{\bf a}$ is an arbitrary unit vector.
 
The soft variables $\phi$ and $R_{\alpha i}$ are determined by the
interaction of the order parameter with various perturbations. The
perturbations can be divided into external fields and boundary
conditions. Experimentally, the most common field is the magnetic one
${\bf H}$. It would also be straightforward to include the electric
field, but we will neglect it here because its coupling is very weak
\cite{SEP}. The motion of the $^3$He container can also be treated as
an external field. In equilibrium the normal fluid component (velocity
${\bf v}_{\rm n}$) will follow the motion of the container, and the
only allowed motions are uniform translation and rotation. The former
is automatically taken into account because of Galilean invariance of
the theory. The rotation will appear as a field
$\nabla\times{\bf v}_{\rm n}$, which equals twice the angular velocity.

It is possible to construct a hydrostatic theory for any magnitude of
the external fields. Here we assume the fields are small enough that
the order parameter is not distorted strongly from the bulk form
(\ref{e.2.1}). Strong perturbations, such as a surface or a vortex
core, are treated as boundary conditions. These cause the order
parameter to deviate substantially from the bulk form within a length
scale of the coherence length
$\xi_0\approx 10$ nm, but on longer length scales also they act as weak
perturbations.
 
The degrees of freedom $\phi$ and
$R_{\alpha i}$ differ crucially in the following respect. The mass flow
can be written as
${\bf j}_{{\rm s}}=\tensor{\rho}_{\rm s}\cdot{\bf v}_{{\rm s}}$, where
$\tensor{\rho}_{\rm s}$ is a phenomenological tensor. In the absence of
external fields, $\tensor{\rho}_{\rm s}$ must be a scalar because of
the isotropy of the unperturbed $A_{\alpha i}$ (\ref{e.2.1}). The mass
current has also to be conserved: $\nabla\cdot{\bf j}_{\rm s}=0$. Thus
we arrive at the Laplace equation
$\nabla^2\phi=0$. Adding the boundary conditions, this completely
determines
$\phi$. The external fields cause only a small correction to this.
Throughout the rest of this paper we neglect the small correction and
start out from the assumption that the Laplace equation for
$\phi$ is already solved, and thus ${\bf v}_{\rm s}$ is known.
 
The problem that remains is to determine the rotation matrix $R_{\alpha
i}$. What makes this different from $\phi$ is that there exists
interaction between the nuclear dipole moments of the $^3$He atoms. It
is of the form \cite{LeggettAnn}
\begin{equation} F_{\rm D}=\lambda_{\rm D}\int
d^3r(R_{ii}R_{jj}+R_{ij}R_{ji})=4\lambda_{\rm D}\int
d^3r\cos\theta(1+2\cos\theta).
\label{e.dipgen}\end{equation} Although this interaction is weak, it
partly removes the degeneracy with respect to the rotation matrix. This
means that the spin current is not conserved, but decays on a scale
$\xi_{\rm D}\approx 10\ \mu$m. On the same scale, the rotation angle
$\theta$ becomes fixed to $\arccos(-1/4)\approx 104^\circ$, which
corresponds to the minimum of
$F_{\rm D}$ (\ref{e.dipgen}), but the degeneracy with
respect to the rotation axis
$\hat{\bf n}$ remains. Because no conservation laws exist, the rotation
axis $\hat{\bf n}$ is more susceptible to all kinds of perturbations
than $\phi$. The subject of the rest of this paper is to study the
texture, i.e.\ $\hat{\bf n}({\bf r})$ on a length scale $\gg\xi_{\rm
D}$.
 
We write down the free energy functional that governs the texture. The
form of the energy terms is based on symmetry properties alone.  The
functional is valid in the limit of low fields and velocities, small
gradients of the order parameter and weak coupling between the spin and
orbital parts of the order parameter. The last condition is practically
always satisfied because the coupling is due to the dipole-dipole
interaction (\ref{e.dipgen}), which is small compared to the
superfluid condensation energy by factor
$10^{-6}(1-T/T_c)^{-1}$. We neglect all constant terms, i.e. terms that
do not depend on $\hat{\bf n}$. The leading terms in the expansion can
be written as follows
\begin{equation} F_{\rm DH}=-a\int d^3r(\hat{\bf n}\cdot{\bf H})^2
\label{e.2.5}\end{equation}
\begin{equation} F_{\rm DV}=-\lambda_{\rm DV}\int d^3r[\hat{\bf
n}\cdot({\bf v}_{\rm s} -{\bf v}_{\rm n})]^2
\label{e.2.6}\end{equation}
\begin{equation} F_{\rm HV}=-\lambda_{\rm HV}\int d^3r[{\bf
H}\cdot\tensor R\cdot ({\bf v}_{{\rm s}}-{\bf v}_{{\rm n}})]^2
\label{e.2.7}\end{equation}
\begin{equation} F_{\rm HV1}=-\lambda_{\rm HV1}\int d^3r{\bf H}\cdot
\tensor R\cdot\mbox{\boldmath$\nabla$}\times {\bf v}_{{\rm n}}
\label{e.2.8}\end{equation}
\begin{equation} F_{\rm G}=\int d^3r\left[\lambda_{\rm G1}{\partial
R_{\alpha i}\over\partial r_i}{\partial R_{\alpha j}\over\partial
r_j}+\lambda_{\rm G2} {\partial R_{\alpha j}\over\partial r_i}
{\partial R_{\alpha j}\over\partial r_i}\right].
\label{e.2.9}\end{equation}  
 The dipole-field term $F_{\rm DH}$ is discussed in Refs.\
\cite{EBA,SBE}, the gradient term $F_{\rm G}$ in Refs.\
\cite{BSOB,SBE}, the dipole-velocity
$F_{\rm DV}$ and field-velocity $F_{\rm HV}$ in Ref.\ \cite{BC},
and the first-order field-velocity term 
$F_{\rm HV1}$ in Ref.\ \cite{VolovikMineev}. Because of Galilean
invariance only the combination ${\bf v}_{\rm s}-{\bf v}_{\rm n}$
appears in Equations (\ref{e.2.6}) and (\ref{e.2.7}). The superfluid
velocity does not appear in the gyromagnetic term (\ref{e.2.8}) because
${\bf v}_{\rm s}$ is curl free. Terms that are linear in
$\hat{\bf n}\cdot{\bf H}$,
$\hat{\bf n}\cdot({\bf v}_{\rm s}-{\bf v}_{\rm n})$ or ${\bf
H}\cdot\tensor R\cdot({\bf v}_{{\rm s}}-{\bf v}_{{\rm n}})$ are
prohibited by parity and time-reversal symmetry. Equations
(\ref{e.2.5})-(\ref{e.2.9}) serve as definitions of the parameters $a$,
$\lambda_{\rm DV}$,
$\lambda_{\rm HV}$, $\lambda_{\rm HV1}$,
$\lambda_{\rm G1}$ and $\lambda_{\rm G2}$, which depend on temperature
and pressure. The calculation of these parameters are discussed in
Sections
\ref{s.qc} and
\ref{s.gl}. For some of the parameters we use names given in Ref.\
\cite{SBE} instead of the more systematic names introduced here; for
example
$\lambda_{\rm DH}\equiv a$ and $\lambda_{\rm SH}\equiv d$.  
 
The derivation of the gradient energy (\ref{e.2.9}) deserves special
consideration. Originally, one starts from a general expression that is
quadratic in the spin velocity (\ref{e.2.3}). Making use of the
properties of the rotation matrices,
it is possible to simplify the energy to a form that
is bilinear in the rotation matrices $R_{\alpha i}$. 
In addition to the two terms in (\ref{e.2.9}), this form contains a
third  term of the form $\partial_iR_{\alpha j}\partial_jR_{\alpha
i}$.    The form (\ref{e.2.9}) can then be obtained by partial
integration which converts the third term into the form 
$\partial_iR_{\alpha i}\partial_jR_{\alpha j}$.

It should be noted that the partial integration of the gradient energy
(\ref{e.2.9})  produces a surface term that is similar to $F_{\rm SG}$
below. Thus the value of the surface coefficient
$\lambda_{\rm SG}$ is unique only if the form of the bulk gradient 
energy is properly defined. Here the uniqueness of $\lambda_{\rm SG}$ is
guaranteed by restricting the bulk gradient energy to the form
(\ref{e.2.9}).
  
The gradient term can also be expressed explicitly as a function
of
$\hat{\bf n}$ using the representation (\ref{e.2.4}) \cite{SBE}. The
needed identities are given in the Appendix. Our preference is to keep
the shorter form (\ref{e.2.9}) because a numerical algorithm can
directly be based on it.

The dipole length $\xi_{\rm D}$ is defined by $\xi_{\rm
D}=\sqrt{\lambda_{\rm G2}/\lambda_{\rm D}}$. It is conventional to
define dipole velocity
$v_{\rm D}$ and dipole field $H_{\rm D}$ by writing $\lambda_{\rm
HV}=2a/(5v_{\rm D}^2)$ and
$\lambda_{\rm DV}=aH_{\rm D}^2v_{\rm D}^{-2}$. We can also define a
magnetic coherence length
$\xi_{\rm H}=\sqrt{65\lambda_{\rm G2}/(8aH^2)}$, which is inversely
proportional to the field. The parameters defined here are temperature
dependent. Near $T_{\rm c}$ they reduce to constants that are commonly
used. For example, $\xi_{\rm H}\rightarrow R_{\rm c}H_{\rm B}/H$
defined in Ref. \cite{SBE}.
 
In addition to the bulk terms (\ref{e.2.5})-(\ref{e.2.9}), there are
boundary terms. These energy terms originate from regions where the
order parameter is strongly distorted from the form (\ref{e.2.1}). We
are here interested in two cases: surfaces and vortex cores. The
boundary terms below are valid in the limit that the length scale of
the distorted region ($\approx\xi_0$) is small compared to the dipole
length $\xi_{\rm D}$. In reality this is well the case. It guarantees
that the rotation angle $\theta$ is not affected by the boundary. The
form of the allowed boundary terms depends on the symmetry of the order
parameter at the boundary.

We assume that the surface structure has the maximal symmetry, i.e.\
time-inversion symmetry, rotation symmetry around the surface normal and
reflection symmetry in planes perpendicular to the surface. (We note
that also less symmetric states are possible \cite{T86}.)  We also
assume that the curvature of the surface is small. Such a surface gives
rise to the energy terms 
\begin{equation} F_{\rm SH}=-d\int_S d^2r({\bf H}\cdot\tensor R\cdot\hat
{\bf s})^2
\label{e.2.10}\end{equation}
\begin{equation} F_{\rm SHV1}=-\lambda_{\rm SHV1}
\int_Sd^2r{\bf H}\cdot\tensor R\cdot\hat {\bf s}\times({\bf v}_{{\rm
s}}-{\bf v}_{{\rm n}})
\label{e.2.11}\end{equation}
\begin{equation} F_{\rm SG}=\lambda_{\rm SG}\int_Sd^2r\hat s_jR_{\alpha
j}{\partial R_{\alpha i}\over\partial r_i}
\label{e.2.12}\end{equation}
\begin{equation} F_{\rm SD}=\int_Sd^2r[ b_4(\hat{\bf s}\cdot\hat{\bf
n})^4-b_2(\hat{\bf s}\cdot\hat{\bf n})^2].
\label{e.2.13}\end{equation}  Here
$\hat{\bf s}$ is a unit vector that is perpendicular to the surface and
points towards the superfluid. The surface-field term $F_{\rm SH}$ is
discussed in Ref.\
\cite{BSOB,SBE}, the surface-dipole term $F_{\rm SD}$ in Ref.\
\cite{BSOB,FominVuorio,SBE}, and the first-order surface-field-velocity
term
$F_{\rm SHV1}$ in Ref.\
\cite{VolovikMineev}. There are two contributions to the
surface-gradient coefficient, 
$\lambda_{\rm SG}=\lambda_{\rm SG}^{\rm a}+\lambda_{\rm SG}^{\rm
b}$. The former comes from the equilibrium spin current that flows
spontaneously along any surface\cite{ZKT}. In fact, the surface spin
current 
$J^{\rm ss}_{\alpha i}=\lambda_{\rm SG}^{\rm a}R_{\alpha
j}\epsilon_{ijk}\hat s_k$. (Note that $J^{\rm ss}$ contains the
factor 
$\hbar/2$ for each fermion and thus has the unit J/m.) The other
contribution
$\lambda_{\rm SG}^{\rm b}$ comes from the partial integration that
depends on the chosen form of
$F_{\rm G}$ \cite{SBE}. Note that there exists only one
surface-gradient term (\ref{e.2.12}) because
$R_{\alpha i}\nabla R_{\alpha j}$ is antisymmetric in $i$ and $j$. For
the same reason the term (\ref{e.2.12}) does not depend on the normal
derivative. We have constructed the definitions
(\ref{e.2.5})-(\ref{e.2.13}) so that all surface ($d$, $\lambda_{\rm
SHV1}$, $\lambda_{\rm SG}$,
$b_2$,
$b_4$)  and bulk coefficients  are non-negative, at least in the
Ginzburg-Landau region.

The order parameter is strongly distorted from the bulk form
(\ref{e.2.1}) in the cores of quantized vortex lines. Therefore the
cores must be treated as boundary regions. We describe a vortex line by
unit vector $\hat{\bf l}$ that is parallel to the line and points in
the direction of the circulation $\nabla\times{\bf v}_{\rm s}$. The
maximal point-symmetry operations of a vortex are  a) reflection in
plane perpendicular to $\hat{\bf l}$, b) rotation around $\hat{\bf l}$
(combined with a phase shift) and c) reflection in plane containing
$\hat{\bf l}$. The last one has to be combined with time inversion
because otherwise the circulation would change direction. Assuming that
the order parameter in the core has all these symmetries, we get the
phenomenological terms
\begin{equation} F_{\rm LH}=\lambda_{\rm LH}
\int_L d^3r({\bf H}\cdot\tensor R\cdot\hat {\bf l})^2
\label{e.2.14}\end{equation}
\begin{equation} F_{\rm LH1}=\lambda_{\rm LH1}\int_Ld^3r{\bf
H}\cdot\tensor R\cdot\hat {\bf l}
\label{e.2.15}\end{equation}
\begin{equation} F_{\rm LD}=\lambda_{\rm LD}\int_Ld^3r[(\hat{\bf
l}\cdot\hat{\bf n})^2+\hbox{corrections}].
\label{e.2.16}\end{equation}
 The line-field term $F_{\rm LH}$ is discussed in Ref.\
\cite{GGK}, the first-order line-field term $F_{\rm LH1}$ in Ref.\
\cite{HKSSBMV}, and line-dipole term
$F_{\rm LD}$ in Ref.\ \cite{T87}.  Here $L$ denotes the region where
vortices are present. In $F_{\rm LD}$ only the dominant term is written
explicitly. 

It is well known that the vortex cores do not have the maximal symmetry
\cite{T87}. In the A-phase-core vortex the symmetry (a) is broken.
Because this can take place in two different ways, we have to assign to
each vortex line a new variable
$q$ that equals either +1 (left-handed vortex) or -1 (right handed
vortex). This allows the line-gradient term \cite{SVrev}
\begin{equation} F_{\rm LG}=\lambda_{\rm LG}\int_Ld^3r\, \langle
q\rangle \hat l_jR_{\alpha j}{\partial R_{\alpha i}\over\partial r_i}, 
\label{e.2.17}\end{equation}
where $\langle \ldots\rangle$ denotes the average because $q$ may
change from one vortex to another. Similar to the surface term
$F_{\rm SG}$,
$F_{\rm LG}$ arises from spontaneous spin currents. For an
isolated vortex these currents form closed loops in the plane
perpendicular to
$\hat{\bf l}$. All vortices in $^3$He-B also have axial
spin currents but they do not
couple to external spin velocity in the lowest order because the net
current vanishes.

The
double-core vortex also allows the term (\ref{e.2.17}). Additionally,
the circular symmetry (b) is broken leaving only discrete symmetry in
rotations by
$\pi$. Thus an additional unit vector
$\hat{\bf b}$ perpendicular to $\hat{\bf l}$ is needed to describe the
vortex. This gives rise to line-anisotropy terms
\begin{equation} F_{\rm LAH}=\lambda_{\rm LAH}
\int_L d^3r\langle({\bf H}\cdot\tensor R\cdot\hat {\bf b})^2\rangle
\label{e.2.18}\end{equation}
\begin{equation} F_{\rm LAD}=\lambda_{\rm
LAD}\int_Ld^3r\langle(\hat{\bf b}\cdot\hat{\bf
n})^2+\hbox{corrections}\rangle.
\label{e.2.19}\end{equation}
 
We note that there is flexibility in the definitions of the different
terms. For example, the superflow around a vortex has to be counted
into term $F_{\rm LH}$ (\ref{e.2.14}) in the region where the order
parameter is strongly distorted but at larger distances it also can be
included as the bulk term $F_{\rm HV}$ (\ref{e.2.7}). 

\section{Connection to the quasiclassical theory}\label{s.qc}

The energy terms (\ref{e.dipgen})-(\ref{e.2.19}) contain a number of
phenomenological coefficients. They should either be determined
experimentally or calculated from a more microscopic theory than the
hydrodynamic one. Pursuing the latter, there exists the quasiclassical
theory \cite{SRrev}. This theory bypasses the difficult many-body
problem of strongly interacting $^3$He
atoms by concentrating in the low-energy range. It uses an
expansion, where the relevant expansion parameter for the
superfluid phases is the transition temperature divided by the Fermi
temperature, $T_{\rm c}/T_{\rm F}\sim 0.001$. The lowest nontrivial
order in this expansion is known as the {\it weak-coupling} theory. It
effectively contains the Bardeen-Cooper-Schrieffer theory as a special
case, but it also reduces to the Landau Fermi-liquid theory in the
normal state. This theory is adequate for some properties of superfluid
$^3$He, especially at low pressures, but it fails, for example, to
stabilize the A phase. For many purposes it is important to continue
the expansion to the next order in $T_{\rm c}/T_{\rm F}$.
We call this the {\it strong-coupling}
theory. (Serene and Rainer use the name ``weak-coupling plus'', but we
think this is too modest since there seems to be very little hope to
calculate further orders in the expansion.)
We will not go into the details of the quasiclassical
theory, which is extensively discussed by Serene and Rainer
\cite{SRrev}. 

It is important to realize that the 
quasiclassical theory is not microscopic in the sense that it would
depend only on fundamental constants. Instead, it needs several
parameters as input. This is especially a problem in the
strong coupling case, which needs as input the scattering amplitude of
quasiparticles (in the normal state) that is not accurately known.
Additionally, the needed calculations are rather complicated at general
temperature. There are two practical ways to proceed. The first is to
restrict to the temperature region close to
$T_{\rm c}$ and use the Ginzburg-Landau theory. This approach will
be described in Section \ref{s.gl}. The second way is to work at
arbitrary temperature but to use the weak-coupling approximation in the
quasiclassical theory. The latter approach is discussed in this section.
At the end of this section we discuss how to improve the weak-coupling
results by including a trivial strong-coupling correction.
 
In the weak-coupling theory, the properties of the normal state are
included via spin symmetric and antisymmetric Fermi-liquid
parameters, 
$F_l^{\rm s}$ and
$F_l^{\rm a}$ ($l=0$, 1, 2, $\ldots\infty$). We assume that the pairing
interaction is effective in the
p-wave channel only. The symmetry between particle and hole types of
quasiparticles is consistently assumed in the quasiclassical theory. It
turns out that all results presented below depend only on five
Fermi-liquid parameters: 
$F_1^{\rm s}$ and
$F_l^{\rm a}$ with $l=0$, 1, 2, and 3. (Infinite number of coefficients
is  needed in the hydrostatics of the A phase \cite{SRres}.) In
addition, the results depend on  the mass density $\rho$ of $^3$He
liquid, on the superfluid transition temperature $T_{\rm c}$ and on the
magnetic dipole-dipole interaction parameter $g_{\rm D}$. 

In the Bardeen-Cooper-Schrieffer model $g_{\rm D}$ has the expression 
\cite{LeggettAnn} (in SI units)
\begin{equation} g_{\rm D}={\mu_0\over 40}\bar{R^2}\left(\hbar\gamma
N(0)\pi k_{\rm B}T\sum_{\epsilon_n=-\epsilon_{\rm c}}^{\epsilon_{\rm c}}
{1\over\sqrt{\epsilon_n^2+\Delta^2}}\right)^2,
\label{e.gd}\end{equation} where $\bar{R^2}$ is a renormalization
constant and $\epsilon_{\rm c}$ a high energy cut-off. (Note that our
definition of $g_{\rm D}$ \cite{Fetterrev} is different from that in
Ref.\ \cite{LeggettAnn}.) The
Matsubara energies $\epsilon_n$ and the weak-coupling
energy gap  $\Delta(T)$ are defined in the Appendix, the
gyromagnetic ratio of the
$^3$He nucleus $\gamma=-2.04\cdot10^8\ ({\rm T\ s})^{-1}$, and the
density of states at the Fermi energy $2N(0)=(1+{1\over 3}F_1^{\rm
s})(3m^2\rho/\pi^4\hbar^6)^{1/3}$. It is very convenient that the
dependence of
$g_{\rm D}$ on temperature is so weak that we can safely ignore it. In
the weak-coupling approximation the constancy of $g_{\rm D}(T)$ would
be exact if the cut-off energy $\epsilon_{\rm c}$ in (\ref{e.gd}) were
the same in the gap equation (\ref{e.qcgap}). (For the standard choice
$\epsilon_{\rm c}\rightarrow\infty$ in the gap equation and
$\bar{R^2}=1$, the relative variation of
$g_{\rm D}(T)$ is less than
$10^{-5}$.) In the trivial strong-coupling model (see below) Eq.
(\ref{e.gd}) gives the maximum variation at the melting pressure, where
$g_{\rm D}$ decreases monotonically by 1.3\%  when $T$ decreases from
$T_{\rm c}$ to zero (assuming
$\bar{R^2}=1$). Because of uncertainties associated with $\epsilon_{\rm
c}$ and
$\bar{R^2}$ in Eq.\ (\ref{e.gd}), we prefer to extract $g_{\rm D}$ from
experiments, as will be discussed in section \ref{s.gd}. 

For completeness, we give the results for nuclear magnetic
susceptibility $\chi$
\cite{SRQ77}, superfluid density $\rho_{\rm s}$, and $\lambda_{\rm D}$
(\ref{e.dipgen})
\begin{equation} \chi=2\mu_0N(0)\left({\hbar\gamma\over
2}\right)^2{{2\over 3}+({1\over 3}+{1\over 5}F^{\rm a}_2)Y\over
1+F_0^{\rm a}({2\over 3}+{1\over 3}Y)+{1\over 5}F_2^{\rm a} ({1\over
3}+({2\over 3}+F_0^{\rm a})Y)}
\label{e.suskis}\end{equation}
\begin{equation} \rho_{\rm s}=\rho{1-Y\over 1+{1\over 3}F_1^{\rm s}Y}
\label{e.rhos}\end{equation}
\begin{equation} \lambda_{\rm D}= g_{\rm D}\Delta^2.
\label{e.lambdad}\end{equation} All the following coefficients can be
understood as corrections to these. Here $Y(T)=1-Z_3(T)$ is the Yoshida
function, and the functions
$Z_j(T)$ are defined in the Appendix. 
 
The basic principle for calculating the hydrostatic parameters is
explained in Section VI of Ref.\
\cite{SRrev}.  For the coefficient of the dipole-field energy $F_{\rm
DH}$ the main part of the work, the calculation of the gap distortion,
is explained in detail in Ref.\
\cite{FS}. The result is
\begin{equation} a={5g_{\rm D}\over 2}\left[ {{1\over
2}\hbar\gamma\mu_0(1+{1\over 5}F^{\rm a}_2)\over 1+F_0^{\rm a}({2\over
3}+{1\over 3}Y)+{1\over 5}F_2^{\rm a} ({1\over 3}+({2\over 3}+F_0^{\rm
a})Y)}\right]^2 \left[5-{3Z_5\over Z_3} -{3F^{\rm a}_2Z_3\over
5(1+{1\over 5}F^{\rm a}_2)}\right].
\label{e.3.1}\end{equation}   The coefficient of the dipole-velocity
energy (\ref{e.2.6}) can be calculated in a similar way and we obtain 
\begin{equation} \lambda_{\rm DV}=5g_{\rm D}\left({m^*v_{\rm F}\over
1+{1\over 3}F_1^{\rm s}Y}\right)^2 \left(1-{3Z_5\over 2Z_3}\right).
\label{e.3.3}\end{equation} Here $m^*$ is the effective mass given by
$m^*/m=1+F^{\rm s}_1/3$. The Fermi velocity $v_{\rm F}$ is related to
basic parameters by $v_{\rm F}=\hbar(3\pi^2\rho/m)^{1/3}/m^*$. 
As far as we know, the expressions
(\ref{e.3.1}) and (\ref{e.3.3}) have not been published before. Rather
tedious calculation is needed for the coefficient in the field-velocity
energy (\ref{e.2.7}).  This is done in Ref.\ \cite{ghv}, and we quote
the result
\begin{eqnarray} \lambda_{\rm
HV}&=&{\rho\over\Delta^2}{m^*/m\over(1+{1\over 3}F_1^{\rm s}Y)^2}\left[
{{1\over 2}\hbar\gamma\mu_0(1+{1\over 5}F^{\rm a}_2)\over 1+F_0^{\rm
a}({2\over 3}+{1\over 3}Y)+{1\over 5}F_2^{\rm a} ({1\over 3}+({2\over
3}+F_0^{\rm a})Y)}\right]^2\nonumber \\ &&\times\left[Z_3 -{9\over
10}Z_5+{9\over 10}{Z_5^2\over Z_3}-{3\over 2}Z_7+{3F^{\rm a}_2Z_3\over
50(1+{1\over 5}F^{\rm a}_2)}(3Z_5-2Z_3)\right].
\label{e.3.4}\end{eqnarray}   The gyromagnetic coefficient
$\lambda_{\rm HV1}=0$ because of particle-hole symmetry
\cite{Mineev}.   The gradient energy coefficients are calculated in
Ref.\ \cite{Dorfle}, and can also be found in Appendix F of Ref.\
\cite{SRrev}. They are
\begin{equation} \lambda_{\rm G2}={\hbar^2\rho\over 40mm^*} {(1+{1\over
3}F_1^{\rm a})(1+{1\over 7}F_3^{\rm a})(1-Y)\over 1+{1\over 3}F_1^{\rm
a}({2\over 5}+{3\over 5}Y)+{1\over 7}F_3^{\rm a} ({3\over 5}+({2\over
5}+{1\over 3}F_1^{\rm a})Y)}
\label{e.3.5}\end{equation}
\begin{equation} {\lambda_{\rm G1}\over\lambda_{\rm G2}}=2+ {({1\over
3}F_1^{\rm a}-{1\over 7}F_3^{\rm a})(1-Y)\over (1+{1\over 7}F_3^{\rm
a})(1+{1\over 3}F_1^{\rm a}Y)}
\label{e.3.6}\end{equation}
 
The structure of the order parameter near surfaces has been studied for
a long time (for example in Refs.\
\cite{Buchholtz} and \cite{ZKT}), but the surface
terms have been evaluated only quite
recently. The gyromagnetic surface term (\ref{e.2.11}) vanishes
identically because of particle-hole symmetry. As discussed above, the
surface-gradient (\ref{e.2.12}) term has two contributions:
$\lambda_{\rm SG}=\lambda_{\rm SG}^{\rm a}+\lambda_{\rm SG}^{\rm b}$.
For the part that arises from the partial integration in the derivation
of (\ref{e.2.9}) we find 
$\lambda_{\rm SG}^{\rm b}=2\lambda_{\rm G2}$. The other part
$\lambda_{\rm SG}^{\rm a}$ coming from spontaneous spin currents has
recently been calculated in Ref.\
\cite{VT}. The same reference evaluates also the
surface dipole coefficients in (\ref{e.2.13}). The field term
(\ref{e.2.10}) has not been calculated. Until this
is done, we can use an extrapolation of the Ginzburg-Landau result 
$d={\mu_0\over 2}(\chi_{\rm n}-\chi)\xi_{\rm GL}d_0$, where
$d_0=d/g_{\rm H}\Delta^2\xi_{\rm GL}$ is plotted by solid lines in Fig.\
\ref{f.surfgl}. The Ginzburg-Landau coherence length
$\xi_{\rm GL}$ can be extrapolated to general temperature by
$\xi_{\rm GL}(T)=\hbar v_{\rm F}/\sqrt{10}\Delta(T)$. (Note that no
strong-coupling correction to the weak-coupling $\Delta$ is allowed in
this equation.)
$\chi_{\rm n}$ is the susceptibility in the normal
state [given by (\ref{e.suskis}) with $Y=1$].

For accurate calculation of the vortex terms, one needs a calculation
of the order parameter in the vortex core. This has been done at
general temperature by Fogelstr\"om and Kurkij\"arvi
\cite{FogelstromKurkijarvi}, but they do not give explicit values of
$\lambda_{\rm LH}$. However, at not too high rotation velocities, the
most of the contribution to
$\lambda_{\rm LH}$ comes from outside of the vortex core, and therefore
a reasonable estimate at all temperatures is 
\begin{equation} \lambda_{\rm LH}\approx {1\over 2}\lambda_{\rm
HV}\langle\vert{\bf v}_{\rm s}-{\bf v}_{\rm n}\vert^2\rangle_L\approx
{\hbar\over 2m}\Omega\lambda_{\rm HV}(\ln{R\over r_{\rm L}}-{3\over 4}),
\label{e.lambdaestimate}\end{equation} where $\Omega$ is the angular
velocity of rotation, $R=\sqrt{\hbar/2m\Omega}$ the unit cell radius of
a vortex, and $r_{\rm L}$ the radius of the vortex core. Because
$\lambda_{\rm LH}$ is rather insensitive to $r_{\rm L}$, we may use
$r_{\rm L}\approx\xi_{\rm GL}$ \cite{T87}, and use the same
extrapolation of
$\xi_{\rm GL}$ as for the surface term.

The weak coupling approximation used above is not expected to be
accurate at high pressures, where strong coupling corrections are
largest. As mentioned above, accurate strong-coupling calculations are
very cumbersome, and introduce the scattering amplitude, which is
poorly known. However, there is a simple procedure that is expected to
take into account a major part of the strong coupling effects. This
``trivial strong coupling correction'' is to scale the energy gap
$\Delta$ by a temperature and pressure dependent factor. This factor is
tabulated by Serene and Rainer
\cite{SRrev} as a function of the temperature and the relative jump
$\Delta C/C_{\rm n}$ of the specific heat at
$T_{\rm c}$. The latter can be related to pressure according to the
measurements by Greywall
\cite{Greywall}. Such scaling of $\Delta$ affects all hydrostatic
coefficients (\ref{e.gd})-(\ref{e.lambdaestimate}) directly and/or via
modification of the functions $Z_j$ and $Y$.

\section{Connection to the Ginzburg-Landau theory}\label{s.gl}

The hydrostatic theory was based on expansion of the free energy in
small gradients and external fields.  The Ginzburg-Landau
(GL) theory is based on additional expansion in the amplitude of
the order parameter
\cite{GL}. The expansion can be limited to a small number of terms near
the superfluid transition temperature
$T_{\rm c}$, where the order parameter is small. In
most superconductors and in $^3$He, the GL theory gives
reliable results in the neighborhood of $T_{\rm c}$ because the
fluctuation range, where it becomes invalid, consists of a negligible
temperature range just at $T_{\rm c}$.

The order parameter in $^3$He is $3\times 3$ matrix $A_{\alpha
j}$. The GL theory consists of writing down the
terms in the free energy that are allowed by known symmetries. The
superfluid condensation energy must be invariant in separate rotations
in the spin and orbital spaces. This allows the leading terms 
\cite{MerminStare}
\begin{eqnarray}
F_{\rm cond}&=&\int d^3r\big[-\alpha A_{\mu i}^*A_{\mu i}
+\beta_1 A_{\mu i}^*A_{\mu i}^*A_{\nu j}  A_{\nu j}
+\beta_2 A_{\mu i}^*A_{\mu i}  A_{\nu j}^*A_{\nu j}\nonumber \\\mbox{}&&
+\beta_3 A_{\mu i}^*A_{\nu i}^*A_{\nu j}  A_{\mu j}
+\beta_4 A_{\mu i}^*A_{\nu i}  A_{\nu j}^*A_{\mu j}
+\beta_5 A_{\mu i}^*A_{\nu i}  A_{\nu j}  A_{\mu j}^*\big].
\label{e.gl}
\end{eqnarray}
The zero of the coefficient $\alpha=\alpha'(1-T/T_{\rm c})$ defines
the transition temperature $T_{\rm c}$. Other coefficients $\beta_i$
($i=1...5$) as well as $\alpha'$ can be taken as constants in
the expansion of $F_{\rm cond}$ to order $(1-T/T_{\rm c})^2$. In
the presence of nonuniform order parameter one needs the gradient 
energy
\cite{AGR}
\begin{eqnarray}
F_{\rm G}&=&K\int d^3r\left[(\gamma-2\eta)(\partial_iA_{\mu
i})^*\partial_jA_{\mu j} +(\partial_iA_{\mu j})^*\partial_iA_{\mu
j}+(2\eta-1)(\partial_iA_{\mu j})^*\partial_jA_{\mu i}
\right]\nonumber\\
&=&K\int d^3r\left[(\gamma-1)({\partial_i}A_{\mu
i})^*\partial_jA_{\mu j} +(\partial_iA_{\mu j})^*\partial_iA_{\mu
j}-{\rm i}(2\eta-1)\epsilon_{kij}(\mbox{\boldmath$\nabla$}
\times{\bf v}_{\rm n})_kA_{\mu i}^*A_{\mu j} \right]\label{e.GLgradient}
\end{eqnarray}
with the Galilean-invariant derivative $\mbox{\boldmath$\partial$}=
\mbox{\boldmath$\nabla$}+2{\rm i}m{\bf v}_{\rm n}/\hbar$. 
In
addition there are the energy caused by the magnetic field ${\bf H}$
\cite{AmbeM,MerminStare}
\begin{equation}
F_{\rm H}=\int d^3r\left(-{\rm i}g_{\rm H1}
\epsilon_{\kappa\mu\nu}H_\kappa A_{\mu i}^*A_{\nu i}+g_{\rm H}H_\mu
A_{\mu i}^*A_{\nu i}H_\nu
+g_{\rm H}'H^2A_{\mu i}^*A_{\mu i}\right),
\label{e.GLfield}
\end{equation}
and the energy of the magnetic
dipole-dipole interaction \cite{LeggettAnn}  
\begin{equation}
F_{\rm D}=g_{\rm D}\int d^3r(A_{i i}^*A_{j j}+A_{i j}^*A_{j
i}-{\textstyle{2\over 3}}A_{\mu i}^*A_{\mu i}).
\label{e.GLdipole}
\end{equation}
We neglect all terms in the free energy that are independent of
$A_{\alpha j}$.  The 
gradient energy (\ref{e.GLgradient}) is parametrized using two
dimensionless parameters $\gamma$ and
$\eta$, which are related to parameters introduced by Serene and Rainer
\cite{SRres} as $\gamma=K_{\rm
L}/K_{\rm T}$ and $\eta=K_{\rm C}/K_{\rm T}$.
The two different forms (\ref{e.GLgradient}) are equivalent, as can
be verified by partial integration. In contrast to the hydrostatic
case (\ref{e.2.9}), the surface term in the partial integration
vanishes here because of the boundary condition $\hat s_iA_{\mu i}=0$. 

The parameters of the GL theory have been calculated using the
weak-coupling quasiclassical theory, and the results are well known (see
Refs.\ \cite{F75,T87}, for example). There are two alternatives to
incorporate the strong-coupling effects. One is to determine the
coefficients purely experimentally. The
$\beta_i$'s, or at least some combinations of them, have been determined
using experiments in the superfluid phases
\cite{THBG,gshift,HBBG}. The other alternative is to consider the
GL theory as a limiting case of the strong-coupling quasiclassical
theory near $T_{\rm c}$ \cite{RSbeta}. Here the
problem of the poorly-known scattering amplitude is encountered again,
but fortunately there exists model calculations for the most
important coefficients. We give here a short summary of the results.

There is a small correction to $\alpha'$ arising from finite lifetime
of quasiparticles \cite{RSspecific}. There are several suggestions for the
$\beta_i$'s that are based on different theoretical assumptions about
the scattering amplitude and measurements in the {\it normal state} of
$^3$He \cite{SS,Bedell,LevinValls}. Although the strong-coupling
corrections generally are small, they can be quite substantial in some
combinations of $\beta_i$'s. For example,
$\beta_{345}\equiv\beta_3+\beta_4+\beta_5$ may differ 50\% from its
weak-coupling value. The corrections to
$K$, $\gamma$, $\eta$,
$g_{\rm H}$, and $g_{\rm H}'$ are calculated by Serene and Rainer
in Ref.\ \cite{SRres}. They find that $\eta$ is unchanged from its
weak-coupling value 1, but $\gamma$  increases from its weak coupling
value 3 to
$\approx 3.1$ at the melting pressure.
$g_{\rm H}'$ is found to vanish even after strong-coupling corrections,
and therefore it is dropped in the following. 
The parameter
$g_{\rm H1}$ vanishes in the quasiclassical theory because of
particle-hole symmetry, but this term is still kept because it
is important in several situations. Its value is best extracted from
measurements of the splitting of the A transition
in magnetic field \cite{Sagan,Israelsson}.
We have assumed that the nontrivial corrections to the dipole energy
(\ref{e.GLdipole}) are small, and therefore use the same coefficient
$g_{\rm D}$ as already discussed in Sect.\ \ref{s.qc}.   

The calculations in the GL theory are considerably simpler than in the
general quasiclassical theory. Essentially all the hydrostatic
parameters appearing in equations (\ref{e.2.5})-(\ref{e.2.19}) have been
calculated. We list below the bulk hydrostatic coefficients as
functions of GL parameters.
\begin{equation} a={5g_{\rm D}g_{\rm H}\over 4\beta_{345}}
\label{e.4.1}\end{equation}
\begin{equation} \lambda_{\rm DV}={5m^2g_{\rm D}(\gamma-1)K
\over\hbar^2\beta_{345}}
\label{e.4.2}\end{equation}
\begin{equation} \lambda_{\rm HV}={2m^2g_{\rm H}(\gamma-1)K
\over\hbar^2\beta_{345}}
\label{e.4.3}\end{equation}
\begin{equation} \lambda_{\rm HV1}={mg_{\rm
H1}(\gamma-4\eta+1)K\over\hbar(\beta_4-3\beta_1-\beta_{35})}
\label{e.4.4}\end{equation}
\begin{equation} \lambda_{\rm G2}=K\Delta^2={\alpha K\over
2(3\beta_{12}+\beta_{345})}
\label{e.4.5}\end{equation}
\begin{equation} {\lambda_{\rm G1}\over\lambda_{\rm G2}}=\gamma-1.
\label{e.4.6}\end{equation} 
The first equality in (\ref{e.4.5}) and (\ref{e.4.6}) can be obtained
trivially by substituting the B-phase order parameter (\ref{e.2.1})
into the gradient energy (\ref{e.GLgradient}). The amplitude $\Delta$ of
the order parameter is obtained by substitution into $F_{\rm cond}$
(\ref{e.gl}) and minimization with respect to $\Delta$. Equations
(\ref{e.4.1})-(\ref{e.4.3}) can be obtained by solving the GL equations
in simple cases of axially distorted B phase. For example, the
coefficient (\ref{e.4.1}) can be obtained by first calculating the
anisotropy of the gap due to a magnetic field and then evaluating the
dipole energy for this gap. The gyromagnetic coefficient
$\lambda_{\rm HV1}$ (\ref{e.4.4}) has been calculated by Mineev
\cite{Mineev}. Because of deviation of $\gamma$ from 3, it is
considerably larger than anticipated in Refs.\
\cite{VolovikMineev,Mineev}.
 
Accurate determination of the surface terms requires a self-consistent
solution of the order parameter near a wall. In the absence of fields,
the order parameter
$\tilde A_{\alpha i}$, which  is normalized to unit matrix in the
bulk, has  real components
$\tilde A_{xx}(x)$ and
$\tilde A_{yy}(x)=\tilde A_{zz}(x)$ near a surface located in the $y-z$
plane. The surface coefficients are then obtained by integration: 
\begin{equation} d=g_{\rm H}\Delta^2\xi_{\rm GL}\int_0^\infty
{dx\over\xi_{\rm GL}}(\tilde A_{yy}^2-\tilde A_{xx}^2)
\label{e.d0}\end{equation} 
\begin{equation}
\lambda_{\rm SG}=K\Delta^2\int_0^\infty dx2(\gamma-1)\tilde
A_{yy}{d\tilde A_{xx}\over dx}
\label{e.sg0}\end{equation} 
\begin{equation} b_2=g_{\rm D}\Delta^2\xi_{\rm GL}\int_0^\infty
{dx\over\xi_{\rm GL}} {5\over 4}(\tilde A_{xx}^2-6\tilde A_{xx}\tilde
A_{yy}+5\tilde A_{yy}^2)
\label{e.b2}\end{equation} 
\begin{equation} b_4=g_{\rm D}\Delta^2\xi_{\rm GL}\int_0^\infty
{dx\over\xi_{\rm GL}} {25\over 8}(\tilde A_{yy}-\tilde A_{xx})^2,
\label{e.b4}\end{equation} where $\xi_{\rm GL}=\sqrt{K/\alpha}$.  The
surface term $F_{\rm SHV1}$ can be found by calculating the order
parameter in the presence of phase gradient: $A_{\alpha j}(x,y)=\Delta
R_{\alpha i}\exp(\imagu ky)\tilde A_{ij}(x)$. The coefficient is then
given by
\begin{equation}
\lambda_{\rm SHV1}={2m\over \hbar}g_{\rm H1}\Delta^2\xi_{\rm
GL}^2\int_0^\infty {dx\over
\xi_{\rm GL}^2}\lim_{k\rightarrow 0}{1\over\imagu k}[\tilde A_{xj}\tilde
A_{yj}^*-\tilde A_{yj}\tilde A_{xj}^*].
\label{e.ks0}\end{equation}

\begin{figure}[p]
\begin{center}\leavevmode
\includegraphics[width=0.4\linewidth]{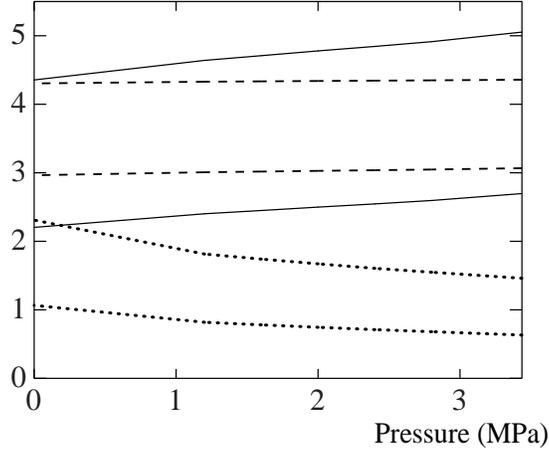}
\caption{  The integrals in Equations (\protect\ref{e.d0}),
(\protect\ref{e.sg0}), and (\protect\ref{e.ks0}). The solid lines give 
$d/g_{\rm H}\Delta^2\xi_{\rm GL}$, dashed  $\lambda_{\rm SG}/K\Delta^2$
and dotted 
$\lambda_{\rm SHV1}\hbar/2mg_{\rm H1}\Delta^2\xi_{\rm GL}^2$. Out of
similar lines, the upper ones always correspond to specular surface
scattering and the lower to diffuse scattering.
}\label{f.surfgl}\end{center}\end{figure}

\begin{figure}[p]
\begin{center}\leavevmode
\includegraphics[width=0.4\linewidth]{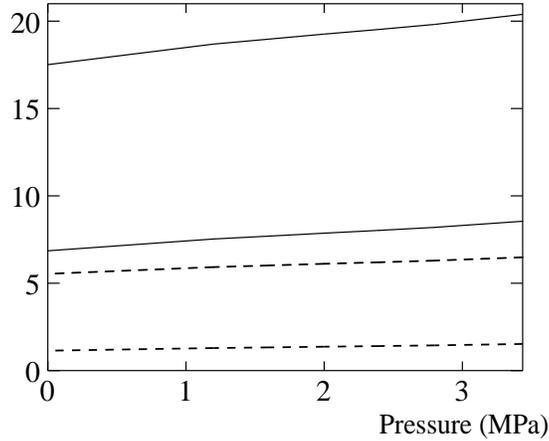}
\caption{  The integrals for surface-dipole terms in Equations
(\protect\ref{e.b2}) and (\protect\ref{e.b4}). The solid lines give 
$b_2/g_{\rm D}\Delta^2\xi_{\rm GL}$ and the dashed $b_4/g_{\rm
D}\Delta^2\xi_{\rm GL}$. For both quantities, the upper lines
correspond to specular surface scattering and the lower to diffuse
scattering.  }\label{f.surfgl2}\end{center}\end{figure}

The surface coefficients $d$, $b_2$, $b_4$ \cite{SBE}, and
$\lambda_{\rm SHV1}$
\cite{VolovikMineev} have been estimated before using  simple models
for the order parameter.  We calculate them here by solving the order
parameter numerically using the Sauls-Serene values
\cite{SS} for the coefficients $\beta_i$ \cite{T86}. For pressures
below 1.2 MPa we smoothly interpolate the parameters to the
weak-coupling values at zero pressure. We also assume the weak-coupling
value
$\gamma=3$. The calculations are done using boundary conditions
appropriate for both specular and diffuse scattering of quasiparticles
\cite{AGR}. (In the latter all the components of the order parameter
have to vanish at the surface.) The integrals in Equations
(\ref{e.d0})-(\ref{e.ks0}) (excluding the prefactors $g_{\rm H}\Delta^2\xi_{\rm GL}$ etc.) are dimensionless, and they are plotted
in Figures
\ref{f.surfgl} and \ref{f.surfgl2}. Because $b_2>2b_4>0$, $F_{\rm SD}$
(\ref{e.2.13}) is minimized by $\hat{\bf n}\parallel\hat{\bf s}$
\cite{FominVuorio}.
 
All the vortex terms except $\lambda_{\rm LG}$ (\ref{e.2.17}) have been
calculated in Ref.\
\cite{T87}.

All the results of this and the previous section are, of course,
identical in the limit where both theories are valid: weak coupling
near $T_{\rm c}$.

\section{Determination of parameters}\label{s.gd}

In his section we analyze a few experiments in order to deduce the
values of parameters $F_0^{\rm a}$, $F_2^{\rm a}$,
and
$g_{\rm D}$. We apply trivial strong-coupling
corrections to the gap
$\Delta$, as explained at the end of Section \ref{s.qc}. For the molar
volume $v=mN_{\rm A}/\rho$ as a function of pressure we assume the fit
in Ref.\ \cite{Greywall}. 

The parameter $F_0^{\rm a}$ can be obtained from measurements in the
normal state: the specific heat and the nuclear susceptibility. For the
specific heat  we use the measurements by Greywall \cite{Greywall}. The
susceptibility has been measured by Ramm et al \cite{Ramm} and by
Hensley et al \cite{Hensley} with essentially identical results.
Unfortunately, it has been measured only below 2.9 MPa, and depending
on the extrapolation alone, the relative error in $1+F_0^{\rm a}$ may
be as large as 10\% at the melting pressure. Examples are the simple
fit $F_0^{\rm a}=-0.909+0.0055 v\ {\rm cm}^{-3}$ and the nonmonotonic
$F_0^{\rm a}(p)$ fit in Ref.\
\cite{HalperinVaroquaux}.

The reduced nuclear susceptibility in the superfluid state,
$\chi(T)/\chi_{\rm n}$ (\ref{e.suskis}), has been measured by
Corruccini and Osheroff \cite{CO}, by Ahonen et al \cite{AKP}, and by
Scholz et al \cite{Hoyt,Scholz}. (There has been a
discrepancy between the susceptibility measured by NMR and by a SQUID
magnetometer 
\cite{HBBG}, but that is probably caused by difficulties in
calibration \cite{Chris}.)
$\chi(T)/\chi_{\rm n}$ depends only on two parameters,
$F_0^{\rm a}$ and $F_2^{\rm a}$. (We treat the trivial strong-coupling
corrections as fixed, and use only the low-field limit of the Scholz
data.) If we assume
$F_0^{\rm a}$ given by Ramm and Hensley et al, only $F_2^{\rm a}$
remains to be fitted. We find that a nonzero {\it pressure-independent}
value of
$F_2^{\rm a}$ does not improve the fit essentially compared to
$F_2^{\rm a}\equiv 0$. Since we believe that $F_2^{\rm a}$ cannot have
strong pressure dependence, the simple choice $F_2^{\rm a}\equiv 0$
seems most attractive to us. Scholz finds $F_2^{\rm a}\approx -1$ with
a weak-coupling fit \cite{Scholz}, but this tendency is largely removed
by the inclusion of trivial strong-coupling corrections. Note that the
susceptibility data could be equally consistent with $F_2^{\rm a}\equiv
-0.7$, say, but that would imply a systematic reduction of $F_0^{\rm
a}$ by -0.025 from the results by Ramm and Hensley et al. Therefore we
take $F_2^{\rm a}\equiv 0$ in the following. For $F_0^{\rm a}$ we use
the simple fit given above because it is in better agreement with the
reduced susceptibility at the melting pressure \cite{CO} than the fit
by Halperin and Varoquaux. Note that at zero pressure our choice is not
far off from the relation between $F_0^{\rm a}$ and
$F_2^{\rm a}$ based on the "catastrophic relaxation" by Bunkov et al
\cite{bunkov}. There exists also other attempts to get $F_2^{\rm
a}$ \cite{FS88,Candela,MKL}.  

The dipole constant $g_{\rm D}$ has to be extracted from experiments
because its value cannot be calculated accurately in the quasiclassical
theory (Sect.\
\ref{s.qc}).   The most straightforward way to get $g_{\rm D}$ is to
measure the B phase longitudinal NMR frequency $\Omega_\parallel$. A
direct measurement of $\Omega_\parallel$ has been made by Bloyet et al
\cite{Bloyet} and  Candela et al \cite{Candela}. These experiments were
done at low temperatures in the collisionless regime. According to the
collisionless theory in a small magnetic field \cite{Fishman}
\begin{equation}
\Omega_\parallel^2={45\Delta^2g_{\rm D}\over\hbar^2N(0)}
\left({1\over\lambda}+{2\over 3}F_0^{\rm a}+{1\over 15}F_2^{\rm
a}\right)
\label{e.omegallcol}\end{equation} where the function $\lambda(T)$ is
defined in the Appendix.
An alternative is to extract
$\Omega_\parallel$ from transverse NMR frequency in surface-oriented
texture. These measurements have been done by Osheroff et al.\
\cite{OEBC}, by Ahonen, Krusius, and Paalanen
\cite{AKP} (results tabulated in Ref.\
\cite{Wheatley}), by Spencer, Alexander and Ihas \cite{Spencer},
and by Hakonen et al. \cite{HKSSSGVK}. Because the external magnetic
field in these experiments reduces the frequency difference between
normal and superfluid precession, these experiments have to be analyzed
using hydrodynamic theory. There $\Omega_\parallel$ is related to
$g_{\rm D}$ via \cite{LeggettAnn}
\begin{equation}
\Omega_\parallel^2={15\mu_0\gamma^2\Delta^2g_{\rm D}\over\chi}.
\label{e.omegall}\end{equation}  Both relations (\ref{e.omegallcol})
and (\ref{e.omegall}) imply that the temperature dependence of
$\Omega_\parallel$ is fully determined by the energy gap
$\Delta$ and the $\lambda$ function (\ref{e.lambda}) or the
susceptibility $\chi$ (\ref{e.suskis}). 

\begin{figure}[p]
\begin{center}\leavevmode
\includegraphics[width=0.5\linewidth]{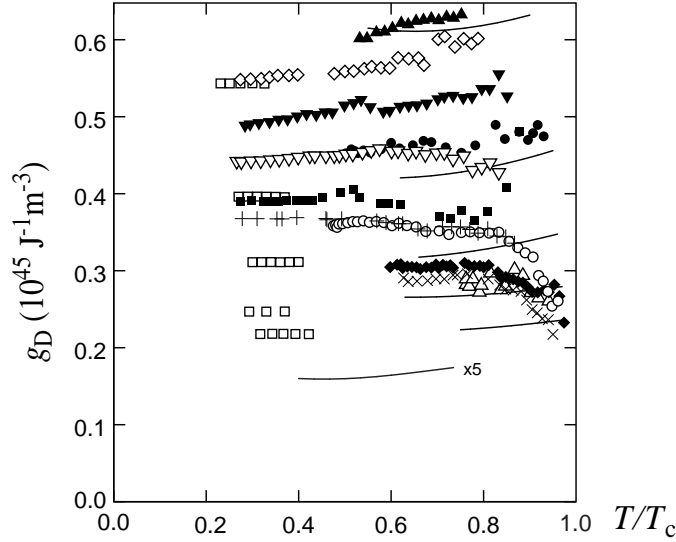}
\caption[GdTemp]{ Plotted as a function of temperature, $g_{\rm D}$
interpreted from different experiments according to trivial
strong-coupling model. In ideal case $g_{\rm D}$ should be independent
of temperature. We have used the transverse NMR frequency measured by
Osheroff et al.\
\protect\cite{OEBC}  ($\blacktriangle$ 3.44 MPa), by Ahonen, Krusius,
and Paalanen
\protect\cite{AKP}  ($\lozenge$ 3.2 MPa, 
$\blacktriangledown$ 2.9 MPa,
$\triangledown$ 2.54 MPa, 
$\blacksquare$ 2.11 MPa 
$+$ 1.87 MPa,), and by Hakonen et al.\ \protect\cite{HKSSSGVK} 
($\bullet$ 2.5 MPa,
$\circ$ 1.55 MPa, 
$\blacklozenge$ 1.02 MPa,
$\times$ 0.5 MPa, 
$\vartriangle$ 0.05 MPa). We also used the longitudinal NMR frequency
measured by Candela et al.\
\protect\cite{Candela} ($\square$, from top to bottom 3.3, 2.1, 1.2,
0.61, and 0.03 MPa).  For comparison, there are curves that are
determined from the g-shift measurements by Osheroff \cite{Osheroff}
(curve with label $\times 5$, reduced by factor ${1/5}$) and by Kycia
et al
\cite{gshift} (curves, from top to bottom 2.17, 1.3, 0.7, 0.3, and 0.11
MPa)   
 }\label{f.GdTemperature}\end{center}\end{figure}

A third way to get $g_{\rm D}$ is so-called g shift of the transverse
NMR frequency $\omega$ from the Larmor frequency $\omega_0$. According
to Ref.\ \cite{SBE}
\begin{equation} {\omega-\omega_0\over\omega_0}={4\mu_0a\over 5\chi}.
\label{e.gshift}\end{equation} The g shift was measured by Osheroff
\cite{Osheroff} at the melting pressure and by Kycia et al
\cite{gshift} below 2.17 MPa. [We note that these measurements are done
in such a high field that the expressions given in this paper are no
more reliable near $T_{\rm c}$. However, Kycia et al measure the
g-shift as a function of magnetization, and this plot is found field
independent, both experimentally and theoretically \cite{gshift}. The
only consequence is that the temperature $T$ in Fig.\
\ref{f.GdTemperature} is the temperature according to the weak-field
susceptibility (\ref{e.suskis}), which differs from the true
temperature near $T_{\rm c}$.]

We plot $g_{\rm D}$ obtained by all three methods in Fig.\
\ref{f.GdTemperature}.  
Let us first ignore the g-shift data (lines). It can be seen that the
data for
$g_{\rm D}$ at each pressure is almost independent of temperature, as
required by theory. We note that in order to reach this constancy it
really is necessary that the longitudinal and transverse data of
$\Omega_\parallel$ are analyzed with collisionless and hydrodynamic
theories, respectively. Equally important is that we use trivial
strong-coupling corrections. The value of $g_{\rm D}$ (but not the
temperature dependence) also depends on
$F_1^{\rm s}$ and $T_{\rm c}$, for which we use the measurements by
Greywall \cite{Greywall}. 

\begin{figure}[p]
\begin{center}\leavevmode
\includegraphics[width=0.45\linewidth]{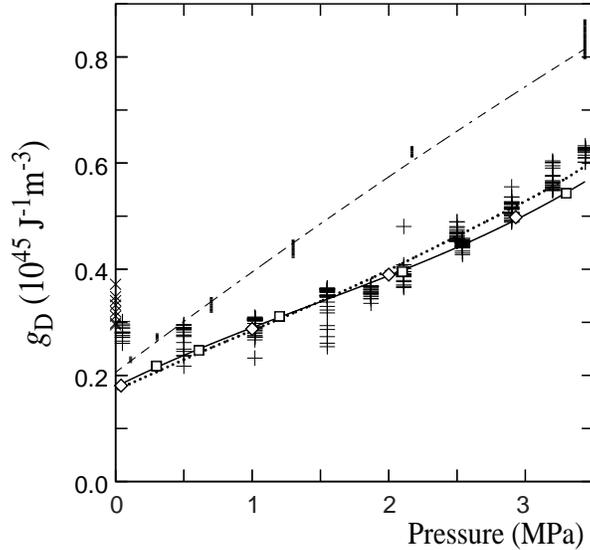}
\caption[GdPres]{ Plotted as a function of pressure, $g_{\rm D}$
interpreted from different experiments according to the trivial
strong-coupling model. The data from $\Omega_\parallel$ in transverse
NMR ($+$), from longitudinal NMR by Candela et al ($\square$), and
g shift (bars) are the same as in Fig.\
\protect\ref{f.GdTemperature}. In addition there is data from zero-field
longitudinal NMR measurements by Bloyet et al \protect\cite{Bloyet}
($\lozenge$) and transverse surface-oriented NMR measurements at
saturated vapor pressure in the temperature range $T/T_{\rm c}=
0.83\ldots 0.86$ by Spencer et al \cite{Spencer} ($\times$). The solid
and dashed lines approximate the data from longitudinal NMR and g shift,
receptively. The dotted line is the model of Eq.\ (\ref{e.gdsimple})
with 
$\epsilon_c=0.45 k_{\rm B}T_{\rm F}$. It is argued in the text that 
$g_{\rm D}$ given by $\Omega_\parallel$ is the correct one, and the
difference between $g_{\rm D}$'s deduced from the g shift and
$\Omega_\parallel$ is a measure of nontrivial strong-coupling 
corrections.
 }\label{f.GdPressure}\end{center}\end{figure}

The $g_{\rm D}$ data as a
function of pressure is plotted in Fig.\ \ref{f.GdPressure}. It
contains all the same data as Fig.\
\ref{f.GdTemperature} and some additional data.  
It can be seen that the longitudinal and transverse
measurements of $\Omega_\parallel$ agree very well at intermediate
pressures, but there is a difference at both high and low pressures. 
Theoretically the ratio of collisionless and hydrodynamic
$\Omega_\parallel$ (at the same pressure) depends only on the reduced
temperatures
$T_1/T_{\rm c}$ and $T_2/T_{\rm c}$ of the two measurements and on
$F_0^{\rm a}$, $F_2^{\rm a}$, and $\Delta C/C_{\rm n}$. The difference
in the effective
$g_{\rm D}$ obtained by the two methods seems at low pressures much
larger than the expected uncertainties of the parameters, and 
remains unexplained. Both the positive slope of $g_{\rm D}(T)$ (Fig.\
\ref{f.GdTemperature}) at high pressures and the difference between
transverse and longitudinal data could be reduced by
giving $F_2^{\rm a}$ a negative value
($\approx -0.7$ at high pressures), but only at the expense of impaired
fit of the $\chi(T)/\chi_{\rm n}$.    

We believe that the nontrivial strong coupling
corrections are small in the expressions for
$\Omega_\parallel$ (\ref{e.omegallcol})-(\ref{e.omegall}), which are
based on expectation values in an unperturbed order parameter. [There
are nontrivial corrections to $\chi$
\cite{SRQ77}, but as long as (\ref{e.suskis}) can reproduce (possibly
with incorrect $F_2^{\rm a}$) the measured $\chi$,  the value obtained
for $g_{\rm D}$ is unaffected.]  The collisionless $\Omega_\parallel$
data is fitted by solid line in Fig.\ \ref{f.GdPressure}. The values
obtained for 
$g_{\rm D}$ depend on $F_0^{\rm a}$, $T_{\rm c}$, $N(0)$, $F_2^{\rm
a}$, and
$\Delta C/C_{\rm n}$, and must be revised if more accurate values of
these become available, for example, via improved measurement of the
temperature \cite{Solen}.

We can also estimate $g_{\rm D}$
based on the simple model of Eq.\ (\ref{e.gd}). Assuming
$\bar{R^2}=1$ and making the sum at
$T=T_{\rm c}$ gives  
\begin{equation} g_{\rm D}={\mu_0\over 40}\left(\hbar\gamma
N(0)\ln{1.1339\epsilon_c\over k_{\rm B}T_{\rm c}}\right)^2.
\label{e.gdsimple}\end{equation}
We take the cut-off energy $\epsilon_c$ proportional to the Fermi
temperature defined by $T_{\rm F}=3\rho/4N(0)k_{\rm B}m$. As shown by
dotted line in Fig.\
\ref{f.GdPressure}, the resulting expression fits nicely the
experimental data for the constant of proportionality 
$\epsilon_c/k_{\rm B}T_{\rm F}=0.45$. The
agreement may be accidental, however, because there is no fundamental
justification for the approximations made. 

Let us next study the $g_{\rm D}$ data based on the g-shift. It is also 
rather independent of the temperature (lines in
Fig.\ \ref{f.GdTemperature}). Fig.\ \ref{f.GdPressure} shows that the
$g_{\rm D}$ data (bars) at different pressures are well consistent with
each other: the low pressure data extrapolates well to the melting
pressure data by Osheroff. However,
$g_{\rm D}$ deduced from the g shift differs essentially from the
determinations based on $\Omega_\parallel$, especially at high
pressures. The reason for this is that the expression for $a$
(\ref{e.3.1}) has substantial strong-coupling corrections that are not
included in the scaling of the energy gap $\Delta$. This can be seen by
comparing the
$T\rightarrow T_{\rm c}$ limit of trivial strong-coupling $a$
(\ref{e.3.1}), denoted by
$a_{\rm TSC}$, with the Ginzburg-Landau limit $a_{\rm GL}$
(\ref{e.4.1}). We find
\begin{equation} {a_{\rm GL}\over a_{\rm TSC}}={a_{\rm GL}\over a_{\rm
WC}}={g_{\rm H}\over g_{\rm H}^{\rm WC}} {\beta_{345}^{\rm WC}\over
\beta_{345}}. 
\label{e.nontrivial}\end{equation} where WC denotes weak-coupling. It
is well known that $\beta_{345}$ differs substantially from its weak
coupling value \cite{SS,THBG,gshift,HBBG}. This explains the difference
in the apparent $g_{\rm D}$ deduced from g shift and $\Omega_\parallel$.
The new thing in the present analysis compared to Ref.\
\cite{gshift} is that the difference is not limited to
the Ginzburg-Landau region, but because of the weak dependency of the
apparent
$g_{\rm D}$ on temperature (Fig.\ \ref{f.GdTemperature}), it persists
almost unchanged at all temperatures. 

\begin{figure}[p]
\begin{center}\leavevmode
\includegraphics[width=0.45\linewidth]{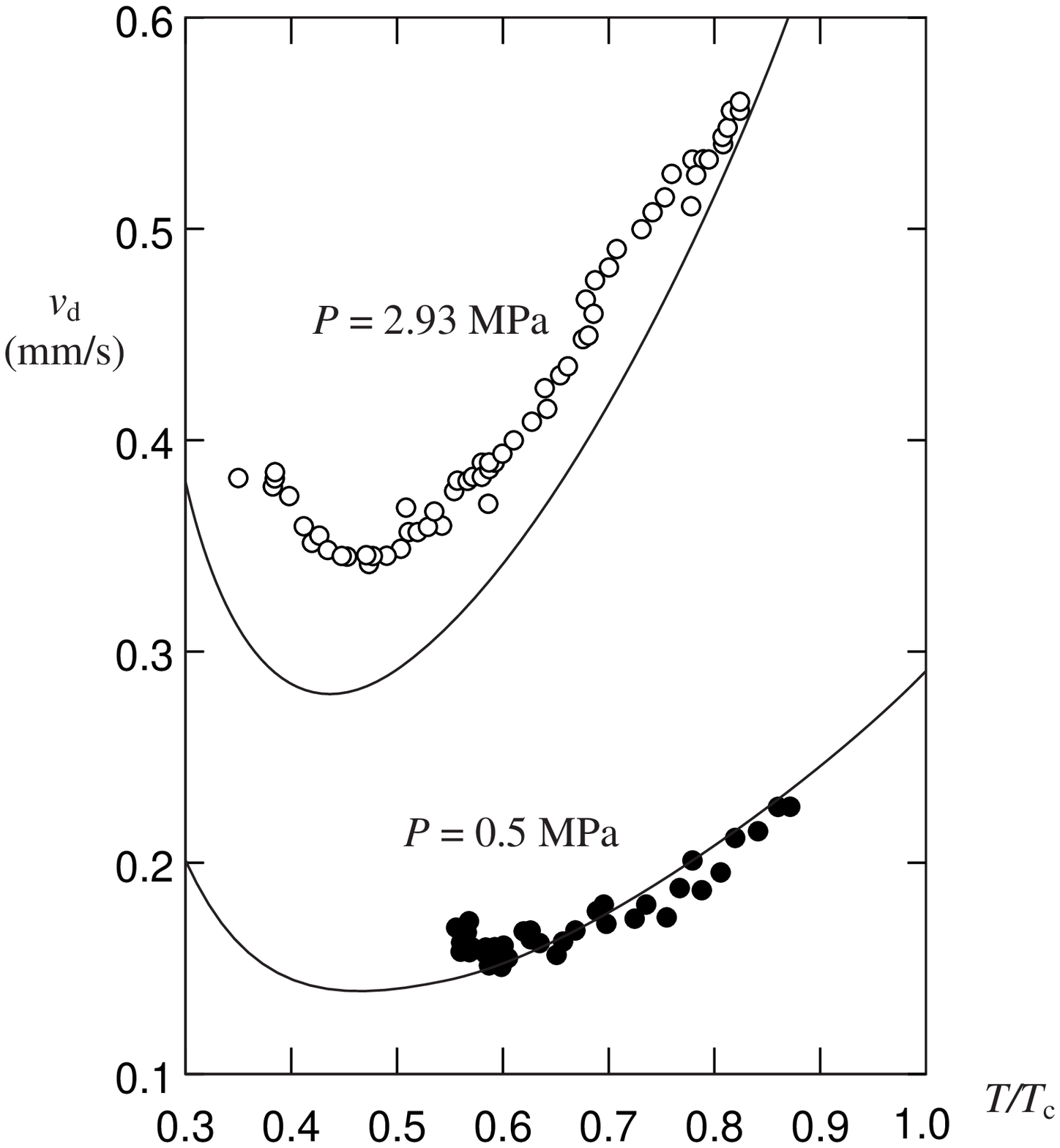}
\caption[vd]{ The dipolar velocity $v_{\rm D}$. The circles are
experimental data from Ref.\ \protect\cite{nummila}. The solid lines are
theory based on Equations (\ref{e.3.1}) and (\ref{e.3.4}), where all
parameter values are fixed by other measurements as explained in the
text. 
 }\label{f.vd}\end{center}\end{figure}

We conclude this section with a comparison
of theoretical and experimental dipole velocity $v_{\rm D}$.
Theoretically this quantity is related to coefficients $a$
(\ref{e.3.1}) and
$\lambda_{\rm HV}$ (\ref{e.3.4}) by the relation
$v_{\rm D}^2=2a/(5\lambda_{\rm HV})$. It has been measured by Nummila
et al \cite{nummila}. Originally they compared their result to a theory
that turned out to be in error, see discussion in Refs.\
\cite{ghv,GKV}. The comparison with the present theory is given in
Fig.\ \ref{f.vd}. We have used trivial strong-coupling theory,
parameters as described above and 
$g_{\rm D}$ from solid line in Fig.\ \ref{f.GdPressure}.  

Because both $a$ (\ref{e.4.1})
and $\lambda_{\rm HV}$ (\ref{e.4.3}) are proportional to
$\beta_{345}^{-1}$ in the Ginzburg-Landau region, the uncertainty
discussed in connection with $a$ is expected to cancel out in $v_{\rm
D}$. There also exists a direct measurement of $\lambda_{\rm HV}$
\cite{ghv}. It also shows deviation from the trivial strong-coupling
model, but the differences are not of similar type as for $a$, and are
presently not understood.

Above we have discussed all input parameters of the
trivial-strong-coupling hydrostatic theory except $F^{\rm a}_1$ and
$F^{\rm a}_3$. Out of the bulk terms only the gradient coefficients
(\ref{e.3.5}) and (\ref{e.3.6})
depend on these. There is several independent evidence that $F^{\rm
a}_1\approx -1$ at high pressures \cite{ORSB,Greywall83,VSexp} but
$F^{\rm a}_3$ is not known. 
 
\section{Conclusions}\label{s.exp}

We have presented a  summary of the hydrostatic theory in superfluid
$^3$He-B. Several new analytic and numerical results were included.
Some experimental data was analyzed in order to extract the
parameters of the theory. A particular goal was to understand how
well the B phase is described by the trivial strong-coupling
model. We found that some quantities ($\chi$,
$g_{\rm D}$, $v_{\rm D}$) can successfully be calculated, but there are
other quantities ($a$, $\lambda_{\rm HV}$) that may be wrong by
50\% in this model. The parameter $g_{\rm D}$ has direct relevance also
for the A phase, where it has been used in comparison between theory and
experiment (see Ref.\ \cite{KT}, for example).   

The first application of the results
calculated here has been the comparison of the ratio
$d/a$ [from equations (\ref{e.d0}) and (\ref{e.4.1})] to experiments in
Ref.\ \cite{KGJKKMT}.  More recently, Kopu et al \cite{kopu} 
applied the hydrostatic theory to a
rotating cylindrical container. They calculated the NMR
line shape and studied the optimal conditions for observing single
vortex lines. Another application is the Josephson $\pi$ state 
observed recently \cite{packard2}. This effect was found to depend
crucially on the texture at the Josephson junction \cite{VTlett,VT}. 
The texture is also essential in several experiments of superfluid 
$^3$He in aerogel. For example, the identification of the B phase was
based on its texture-dependent NMR spectrum \cite{AKWRNH}. For the
present, the textural parameters in aerogel have been evaluated only in
the Ginzburg-Landau region in the homogeneous scattering model
\cite{verditz}. These developments demonstrate that there still are
open problems in superfluid $^3$He and in many cases a proper
understanding of the hydrostatic theory is a prerequisite for
solving them.

\section*{Acknowledgments}

I thank the Academy of Finland for financial support.  
 
\section*{Appendix}

We give here some equations that complete the theory presented above.
The weak-coupling energy gap $\Delta(T)$ is determined by the equation
\begin{equation} 
\ln{T\over T_{\rm c}}+\pi
T\sum_{n=-\infty}^\infty\left[{1\over\vert\epsilon_n\vert}
-{1\over\sqrt{\epsilon_n^2+\Delta^2}}\right]=0,
\label{e.qcgap}\end{equation}
where the
Matsubara energies $\epsilon_n=\pi T(2n-1)$  with $n=0, \pm1,
...\pm\infty$. The $Z_j(T)$ functions are defined by 
\begin{equation} Z_j=\pi k_{\rm B}T\Delta^{j-1}\sum_{n=-\infty}^\infty
(\epsilon_n^2+\Delta^2)^{-j/2},
\label{e.3.2}\end{equation} 
$Y(T)=1-Z_3(T)$. The $\lambda(T)$ function \cite{Fishman}, which
equals to
$1-f(T)$ defined in Ref.\ \cite {LT}, can be written as 
\begin{equation}
\lambda=\pi k_{\rm B}T\sum_{n=-\infty}^\infty
{\Delta\over\sqrt{\epsilon_n^2+\Delta^2}(\sqrt{\epsilon_n^2+\Delta^2}
+\Delta)}.
\label{e.lambda}\end{equation}
The numerical calculation of the functions is discussed in Ref.\
\cite{bcsgap}.

The gradient
energies (\ref{e.2.9}) and (\ref{e.2.12}) can be written in different
forms using the identities \cite{SBE}
\begin{eqnarray} \partial_i R_{\alpha j}\partial_i R_{\alpha j}
&=&4(1-\cos\theta)(\partial_i \hat n_j)^2
=4(1-\cos\theta)\{(\nabla\times\hat{\bf n})^2+(\nabla\cdot\hat{\bf n})^2
 +\nabla\cdot[(\hat{\bf n}\cdot\nabla)\hat{\bf n}
 -\hat{\bf n}(\nabla\cdot\hat{\bf n})]\}\\
\partial_i R_{\alpha i}\partial_j R_{\alpha j}
&=&(1-\cos\theta)[2(\nabla\times\hat{\bf n})^2
 +(1-\cos\theta)(\nabla\cdot\hat{\bf n})^2
 -(1-\cos\theta)(\hat{\bf n}\cdot\nabla\times\hat{\bf n})^2
 -2\sin\theta(\nabla\cdot\hat{\bf n})(\hat{\bf n}\cdot\nabla\times\hat{\bf n})]\\
\partial_i R_{\alpha j}\partial_j R_{\alpha i}
&=&(1-\cos\theta)\{2(\nabla\times\hat{\bf n})^2
 +(1-\cos\theta)(\nabla\cdot\hat{\bf n})^2
 -(1-\cos\theta)(\hat{\bf n}\cdot\nabla\times\hat{\bf n})^2\nonumber\\&&
 -2\sin\theta(\nabla\cdot\hat{\bf n})(\hat{\bf n}\cdot\nabla\times\hat{\bf n})
 +2\nabla\cdot[(\hat{\bf n}\cdot\nabla)\hat{\bf n}
 -\hat{\bf n}(\nabla\cdot\hat{\bf n})]\}\\
\hat s_i R_{\alpha i}\partial_j R_{\alpha j}
&=&-(1-\cos\theta)\hat{\bf s}\cdot[(\hat{\bf n}\cdot\nabla)\hat{\bf n}
 -\hat{\bf n}(\nabla\cdot\hat{\bf n})]\}.
\label{e.drdr}\end{eqnarray} 
Using these it can be seen that $F_{\rm G}$ (\ref{e.2.9}) has a pure
divergence term $\propto\nabla\cdot[(\hat{\bf n}\cdot\nabla)\hat{\bf n}
 -\hat{\bf n}(\nabla\cdot\hat{\bf n})]$. The prefactor of this term is
half of the value of that by Smith, Brinkman and Engelsberg \cite{SBE}.
With present definitions the other half is transferred to the surface
term (\ref{e.2.12}).


\begin{thebibliography}{99}
\bibitem{LounasT} O.V. Lounasmaa and E.V. Thuneberg, Proc. Natl. Acad.
Sci. USA {\bf 96}, 7760 (1999).

\bibitem{aerogel} A. Golov, J.V. Porto, D.A. Geller, N. Mulders, G.J.
Lawes, and J.M. Parpia, Physica B {\bf 280}, 134 (2000).

\bibitem{packard2} A. Marchenkov, R.W. Simmonds, S. Backhaus, A.
Loshak, J.C. Davis, and R.E. Packard, Phys. Rev. Lett. {\bf 83}, 3860
(1999).

\bibitem{LLfluid} L.D. Landau and E.M. Lifshitz, {\it Fluid Mechanics}
(Pergamon Press, Oxford, 1987).

\bibitem{Leggett} A.J. Leggett, Rev. Mod. Phys. {\bf 47}, 331
(1975).
 
\bibitem{BC} W.F. Brinkman and M.C. Cross, in {\it Progress in Low
Temperature Physics, Vol VIIA}, ed. D.F. Brewer (North Holland, 1978),
p. 105.
 
\bibitem{SRrev} J.W. Serene and D. Rainer, Phys. Rep. {\bf 101}, 221
(1983).

\bibitem{Fetterrev} A.L. Fetter, in {\it Progress in Low Temperature
Physics, Vol. X}, ed. D.F. Brewer (Elsevier, 1986), p. 1.

\bibitem{SVrev} M.M. Salomaa and G.E. Volovik, Rev. Mod. Phys. {\bf
59}, 533 (1987), {\bf 60}, 573 (1988).

\bibitem{Kharadze} G.A. Kharadze, in {\it Helium Three}, ed. W.P.
Halperin and L.P. Pitaevskii (Elsevier, Amsterdam 1990), p. 167. 

\bibitem{VW} D. Vollhardt and P. W\"olfle, {\it The superfluid phases
of helium 3} (Taylor \& Francis, London 1990).

\bibitem{BCS} J. Bardeen, L.N. Cooper, and J.R. Schrieffer, Phys. Rev.
{\bf 108}, 1175 (1957).

\bibitem{WheatleyRev} J.C. Wheatley, Rev. Mod.
Phys. {\bf 47}, 415 (1975).

\bibitem{SEP} G.W. Swift, J.P. Eisenstein, and R.E. Packard, Phys. Rev.
Lett. {\bf 45}, 1955 (1980). 

\bibitem{LeggettAnn} A.J. Leggett, Ann. Phys. (New York) {\bf 85}, 11
(1974).

\bibitem{EBA} S. Engelsberg, W.F. Brinkman, and P.W. Anderson, Phys.
Rev. A {\bf 9}, 2592 (1974).

\bibitem{SBE} H. Smith, W.F.
Brinkman, and S. Engelsberg, Phys. Rev. B{\bf 15}, 199 (1977).
 
\bibitem{BSOB} W.F. Brinkman, H. Smith, D.D. Osheroff, and E.I. Blount,
Phys. Rev. Lett {\bf 33}, 624 (1974).

\bibitem{VolovikMineev} G.E. Volovik and V.P. Mineev, Zh. Eksp. Teor.
Fiz. {\bf 86}, 1667 (1984) [Sov. Phys. JETP {\bf 59}, 972 (1984)].
 
\bibitem{T86} E.V. Thuneberg, Phys. Rev. B {\bf 33}, 5124 (1986).
 
\bibitem{FominVuorio} I.A. Fomin and M. Vuorio, J. Low Temp. Phys. {\bf
21}, 271 (1975).

\bibitem{ZKT}W. Zhang, J. Kurkij\"arvi, and E.V. Thuneberg, Phys. Rev.
B{\bf 36}, 1987 (1987).
 
\bibitem{GGK} A.D. Gongadze, G.E. Gurgenishvili, and G.A. Kharadze,
Fiz. Nizk. Temp. {\bf 7}, 821 (1981) [Sov. J. Low Temp. Phys. {\bf 7},
397 (1981)].
 
\bibitem{HKSSBMV} P.J. Hakonen, M. Krusius, M.M. Salomaa, J.T. Simola,
Yu.M. Bunkov, V.P. Mineev, and G.E. Volovik, Phys. Rev. Lett. {\bf 51},
1362 (1983).

\bibitem{T87} E.V. Thuneberg, Phys. Rev. B {\bf 36}, 3583 (1987).
 
\bibitem{SRres} J.W. Serene and D. Rainer, Phys. Rev. B {\bf 17}, 2901
(1978).

\bibitem{SRQ77} J.W. Serene and D. Rainer, in {\it Quantum fluids and
solids}, eds. S.B. Trickey, E.D. Adams, and J.W. Dufty (Plenum, New
York, 1977), p. 111.

\bibitem{FS} R.S. Fishman and J.A. Sauls, Phys. Rev. B {\bf 33}, 6068
(1986).

\bibitem{ghv} J.S. Korhonen, Yu.M. Bunkov, V.V. Dmitriev, Y. Kondo, M.
Krusius, Yu.M. Mukharskiy, \"U. Parts, and E.V. Thuneberg, Phys. Rev. B
{\bf 46}, 13983 (1992).

\bibitem{Mineev} V.P. Mineev, Zh. Eksp. Teor. Fiz. {\bf 90}, 1236
(1986) [Sov. Phys. JETP {\bf 63}, 721 (1986)].
 
\bibitem{Dorfle} M. D\"orfle, Phys. Rev. B {\bf 23}, 3267 (1981); note
incorrect powers of 2 in equations (5.3) and (5.4).
 
\bibitem{Buchholtz} L.J. Buchholtz, Phys. Rev. B {\bf 33}, 1579 (1986).

\bibitem{VT} J. Viljas and E.V. Thuneberg, to be published. 
 
\bibitem{FogelstromKurkijarvi} M. Fogelstr\"om and J. Kurkij\"arvi, J.
Low Temp. Phys. {\bf 98}, 195 (1995); erratum {\bf 100}, 597 (1995).

\bibitem{Greywall} D.S. Greywall, Phys. Rev. B {\bf 33}, 7520 (1986).

\bibitem{GL} V.L. Ginzburg and L.D. Landau, Zh. Eksp. Teor. Fiz. {\bf
20}, 1064 (1950).

\bibitem{MerminStare}  N.D. Mermin and C. Stare, Phys. Rev.
Lett. {\bf 30}, 1135 (1973).

\bibitem{AGR} V. Ambegaokar, P.G. deGennes, and D. Rainer, Phys. Rev. A
{\bf 9}, 2676 (1974), {\bf 12}, 345 (1975).
 
\bibitem{AmbeM} V. Ambegaokar and N.D. Mermin, Phys. Rev. Lett. {\bf
30}, 81 (1973).

\bibitem{F75}  A.L. Fetter, in {\it Quantum Statistics and the
Many-Body Problem}, eds. S.B. Trickey, W.P. Kirk, and J.W. Dufty
(Plenum, New York, 1975), p. 127.

\bibitem{THBG}  Y.H. Tang, I. Hahn, H.M. Bozler, and C.M. Gould, Phys.
Rev. Lett. {\bf 67}, 1775 (1991).

\bibitem{gshift} J.B. Kycia, T.M. Haard, M.R. Rand, H.H. Hensley, G.F.
Moores, Y. Lee, P.J. Hamot, D.T. Sprague, W.P. Halperin, and E.V.
Thuneberg, Phys. Rev. Lett. {\bf 72}, 864 (1994).

\bibitem{HBBG}  I. Hahn, S.T.P. Boyd, H.M. Bozler, and C.M. Gould,
Phys. Rev. Lett. {\bf 81}, 618 (1998).

\bibitem{RSbeta} D. Rainer and J.W.
Serene, Phys. Rev. B {\bf 13}, 4745 (1976).

\bibitem{RSspecific} D. Rainer and J.W. Serene, J. Low Temp. Phys. {\bf
38}, 601 (1980).

\bibitem{SS} J.A. Sauls and J.W. Serene, Phys. Rev. B {\bf 24},
183 (1981).
 
\bibitem{Bedell} K. Bedell, Phys. Rev. B {\bf 26}, 3747
(1982).

\bibitem{LevinValls} K. Levin and O.T. Valls, Phys. Rep. {\bf
98}, 1 (1983).

\bibitem{Sagan} D.C. Sagan, P.G.N. deVegvar, E. Polturak, L. Friedman,
S.-S. Yan, E.L. Ziercher and D.M. Lee, Phys. Rev. Lett. {\bf 53}, 1939
(1984).

\bibitem{Israelsson}  U.E. Israelsson, B.C. Crooker, H.M. Bozler
and C.M. Gould, Phys. Rev. Lett. {\bf 53}, 1943 (1984).

\bibitem{Ramm} H. Ramm, P. Pedroni, J.R. Thompson, and
H. Meyer, J. Low Temp. Phys. {\bf 2}, 539 (1970).

\bibitem{Hensley}  H.H. Hensley, Y. Lee, P. Hamot, T. Mizusaki, and W.P.
Halperin, J. Low Temp. Phys. {\bf 90}, 149 (1993).

\bibitem{HalperinVaroquaux} W.P. Halperin and E. Varoquaux, in {\it
Helium Three}, ed. W.P. Halperin and L.P. Pitaevskii (Elsevier,
Amsterdam 1990), p.\ 353.

\bibitem{CO} L.R. Corruccini and D.D. Osheroff, Phys. Rev. B {\bf 17},
126 (1978).

\bibitem{AKP} A.I. Ahonen, M. Krusius, and M.A. Paalanen, J. Low Temp.
Phys. {\bf 25}, 421 (1976).

\bibitem{Hoyt} R.F. Hoyt, H.N. Scholz, and D.O. Edwards, Physica B {\bf
107}, 287 (1981).

\bibitem{Scholz} H.N. Scholz, Thesis (Ohio State University, 1981). 

\bibitem{Chris} C.M. Gould, private communication.

\bibitem{bunkov} Yu.M. Bunkov,
S.N. Fisher, A.M. Gu\'enault, C.J. Kennedy, and G.R. Pickett, J. Low
Temp. Phys. {\bf 89}, 27 (1992).

\bibitem{FS88} R.S. Fishman and J.A. Sauls, Phys. Rev. B {\bf 38}, 2526
(1988).

\bibitem{Candela} D. Candela, D.O. Edwards, A.
Heff, N. Masuhara, Y. Oda, and D.S. Sherrill, Phys. Rev. Lett. {\bf
61}, 420 (1988). 

\bibitem{MKL} R. Movshovich, N. Kim, and D.M. Lee, Phys. Rev.
Lett. {\bf 64}, 431 (1990).

\bibitem{Bloyet} D. Bloyet, E. Varoquaux, C. Vibet, O. Avenel, P.M.
Berglund, and R. Combescot, Phys. Rev. Lett. {\bf 42}, 1158 (1979); C.
Vibet, thesis, L'Universit\'e Paris-Sud, Centre d'Orsay, 1979
(unpublished).

\bibitem{Fishman} R.S. Fishman, Phys. Rev. B {\bf 36}, 79 (1987).

\bibitem{OEBC} D.D. Osheroff, S. Engelsberg, W.F. Brinkman, and L.R.
Corruccini, Phys. Rev. Lett. {\bf 34}, 190 (1975).

\bibitem{Wheatley} J.C. Wheatley, in {\it Progress in Low Temperature
Physics, Volume VIIA}, ed. D.F. Brewer (North-Holland, 1978), p. 1.

\bibitem{Spencer} G.F. Spencer, P.W. Alexander, and G.G. Ihas, Physica
B {\bf 107}, 289 (1981).

\bibitem{HKSSSGVK} P.J. Hakonen, M. Krusius, M.M. Salomaa, R.H. 
Salmelin, J.T. Simola, A.D. Gongadze, G.E. Vachnadze, and G.A. 
Kharadze, J. Low Temp. Phys. {\bf 76}, 225 (1989). 

\bibitem{Osheroff} D.D. Osheroff, unpublished, g-shift
$\approx (0.8373-1.1677t+1.8736t^2)\cdot 10^{-5}$ for $T/T_{\rm
c}\equiv t$ in the range 0.39-0.74 at the melting pressure.  

\bibitem{Solen} R.J. Solen Jr. and W.E. Fogle, Physics Today {\bf 50},
no. 8 Part 1, 36 (1997).

\bibitem{nummila} K.K. Nummila, P.J. Hakonen, and O.V. Magradze,
Europhys. Lett. {\bf 9}, 355 (1989).

\bibitem{GKV} A. Gongadze, G. Kharadze, and G. Vachnadze, Fiz. Nizk.
Temp. {\bf 23}, 546 (1997) [Low Temp. Phys. {\bf 23}, 405 (1997)].

\bibitem{ORSB} D.D. Osheroff, W. van Roosbroeck, H. Smith, and W.F.
Brinkman, Phys. Rev. Lett. {\bf 38}, 134 (1977).

\bibitem{Greywall83}
D.S. Greywall, Phys. Rev. B {\bf 27}, 2747 (1983).

\bibitem{VSexp} \"U. Parts, V.M.H. Ruutu, J.H. Koivuniemi, M. Krusius,
E.V. Thuneberg, and G.E. Volovik, Physica B {\bf 210}, 311 (1995).

\bibitem{KT} J.M. Karim\"aki and E.V. Thuneberg, Phys. Rev. B {\bf 60},
15290 (1999).

\bibitem{KGJKKMT} J.S. Korhonen, A.D. Gongadze, Z. Jan\'u, Y. 
Kondo, M. Krusius, Yu.M. Mukharsky, and E.V. Thuneberg, Phys.  Rev.
Lett. {\bf 65}, 1211 (1990).

\bibitem{kopu} J. Kopu, R. Schanen, R. Blaauwgeers, V.B.
Eltsov, M. Krusius, J.J. Ruohio, and E.V. Thuneberg, J. Low Temp. Phys.
{\bf 120}, 213 (2000).

\bibitem{VTlett} J.K. Viljas and E.V. Thuneberg, Phys. Rev. Lett. {\bf
83}, 3868 (1999).

\bibitem{AKWRNH} H. Alles, J.J. Kaplinsky, P.S. Wootton, J.D. Reppy,
J.H. Naish, and J.R. Hook, Phys. Rev. Lett. {\bf 83}, 1367
(1999).

\bibitem{verditz} E.V. Thuneberg, in {\it Quasiclassical Methods in
Superconductivity and Superfluidity, Verditz 96}, ed. D. Rainer and
J.A. Sauls (1998) p. 53; cond-mat/9802044.

\bibitem{LT} A.J. Leggett and S. Takagi, Ann. Phys. (New
York) {\bf 106}, 79 (1977).

\bibitem{bcsgap}http://boojum.hut.fi/research/theory/qc/bcsgap.html

\end{thebibliography}
\end{document}